\documentclass[twocolumn]{aastex61}

\newcommand\aastex{AAS\TeX}

\accepted{September 6, 2018}
\submitjournal{ApJ Letters}

\shorttitle{\aastex\ Polarized emission from the CW Tau and DG Tau protoplanetary disks with ALMA}
\shortauthors{F. Bacciotti et al.}
\begin{document}

%%%%%%%%%%%%%%%%%%%%%%%%%%%%%%%%%%%%%%%%%%%%%%%%%%%%%%%%%%%%%%%%%%%%%%%%
%\title{Constraints on grain growth and settling from  ALMA observations of polarized emission of the CW Tau and DG Tau protoplanetary disks} 
%%%%%%%%%%%%%%%%%%%%%%%%%%%%%%%%%%%%%%%%%%%%%%%%%%%%%%%%%%%%%%%%%%%%%%%%
%%%%%%%%%%%%%%%%%%%%%%%%%%%%%%%%%%%%%%%%%%%%%%%%%%%%%%%%%%%%%%%%%%%%%%%%
\title{ALMA observations of polarized emission toward the CW Tau and DG Tau protoplanetary disks: constraints on dust grain growth and settling} 

\correspondingauthor{Francesca Bacciotti}
\email{fran@arcetri.astro.it}

\author[0000-0001-5776-9476]{Francesca Bacciotti}
\affiliation{Istituto Nazionale di Astrofisica - Osservatorio Astrofisico di Arcetri, 
Largo Enrico Fermi, 5, 
I-50125 Firenze, Italy}

\author[0000-0002-3829-5591]{Josep Miquel Girart }
\affiliation{Institut de Ci\`encies de l'Espai (ICE, CSIC), Can Magrans, s/n, E-08193 Cerdanyola del Vallès, Catalonia}
\affiliation{Institut d'Estudis Espacials de Catalunya (IEEC), E-08034, Barcelona, Catalonia}

\author[0000-0003-2303-0096]{Marco Padovani}
\affiliation{Istituto Nazionale di Astrofisica - Osservatorio Astrofisico di Arcetri,
Largo Enrico Fermi, 5, 
I-50125 Firenze, Italy}

\author[0000-0003-2733-5372]{Linda Podio}
\affiliation{Istituto Nazionale di Astrofisica - Osservatorio Astrofisico di Arcetri,
Largo Enrico Fermi, 5,
I-50125 Firenze, Italy}

\author{Rosita Paladino}
\affiliation{Istituto Nazionale di Astrofisica - Istituto di Radioastronomia
Via P. Gobetti, 101 40129 Bologna, Italy}

\author[0000-0003-1859-3070]{Leonardo Testi}
\affiliation{European Southern Observatory, Karl-Schwarzschild-Strasse 2, 85748 
Garching bei M\"unchen, Germany}
\affiliation{Istituto Nazionale di Astrofisica - Osservatorio Astrofisico di Arcetri,
Largo Enrico Fermi, 5,
I-50125 Firenze, Italy}

\author[0000-0001-9249-7082]{Eleonora Bianchi}
\affiliation{Institut de Plan\'etologie et d'Astrophysique de Grenoble (IPAG)
Universit\'e Grenoble Alpes,
CS 40700,
38058 Grenoble C\'edex 9, France}

\author[0000-0001-7706-6049]{Daniele Galli}
\affiliation{Istituto Nazionale di Astrofisica - Osservatorio Astrofisico di Arcetri,
Largo Enrico Fermi, 5,
I-50125 Firenze, Italy}

\author{Claudio Codella }
\affiliation{Istituto Nazionale di Astrofisica - Osservatorio Astrofisico di Arcetri,
Largo Enrico Fermi, 5,
I-50125 Firenze, Italy}

\author[0000-0002-2210-202X]{Deirdre Coffey}
\affiliation{School of Physics, University College Dublin,
Belfield, Dublin 4, Ireland}
\affiliation{School of Cosmic Physics, The Dublin Institute for Advanced Studies, Dublin 2, Ireland}

\author{Cecile Favre}
\affiliation{Istituto Nazionale di Astrofisica - Osservatorio Astrofisico di Arcetri,
Largo Enrico Fermi, 5,
I-50125 Firenze, Italy}

\author[0000-0001-7706-6049]{Davide Fedele}
\affiliation{Istituto Nazionale di Astrofisica - Osservatorio Astrofisico di Arcetri,
Largo Enrico Fermi, 5,
I-50125 Firenze, Italy}

%% Note that the \and command from previous versions of AASTeX is now
%% depreciated in this version as it is no longer necessary. AASTeX 
%% automatically takes care of all commas and "and"s between authors names.

%% AASTeX 6.1 has the new \collaboration and \nocollaboration commands to
%% provide the collaboration status of a group of authors. These commands 
%% can be used either before or after the list of corresponding authors. The
%% argument for \collaboration is the collaboration identifier. Authors are
%% encouraged to surround collaboration identifiers with ()s. The 
%% \nocollaboration command takes no argument and exists to indicate that
%% the nearby authors are not part of surrounding collaborations.

%% Mark off the abstract in the ``abstract'' environment. 

%%%%%%%%%%%%%%%%%%%%%%%%%%%%%%%%%%%%%%%%%%%%%%%%%%%%%%%%%%%%%%%%%%%%%%%%%%
\begin{abstract} 
We present polarimetric data of CW Tau and DG Tau, two well-known Class II disk/jet systems, 
obtained with the Atacama Large Millimeter/submillimeter Array 
at 870 $\mu$m and 0\farcs2 average resolution. 
In CW Tau, the total and polarized emission are both smooth and symmetric,
with polarization angles almost parallel to the minor axis of the projected disk. 
In contrast, DG Tau displays a structured polarized emission, with an elongated brighter region in the disk's near side and a belt-like feature beyond about 0\farcs3 from the source. At the same time the total intensity is 
spatially smooth, with no features. 
The polarization pattern, almost parallel to the minor axis in the inner region, becomes 
azimuthal in the outer belt, possibly because of a drop in optical depth. 
The polarization fraction 
has average values of  1.2\% in CW Tau and  0.4\% in DG Tau. 
Our results are consistent with polarization from self-scattering of the dust thermal emission. 
Under this hypothesis, the maximum size of the grains contributing to polarization
is in the range 100 - 150 $\mu$m for CW Tau and 50 - 70 $\mu$m for DG Tau. 
The polarization maps combined with dust opacity estimates indicate 
that these grains are distributed in a geometrically thin layer in CW Tau, 
representing a settling in the disk midplane. 
Meanwhile, such settling is not yet apparent for DG Tau.
These results advocate polarization studies as a fundamental complement to total emission observations, in investigations of the structure and the evolution of protoplanetary disks. 

\end{abstract} 
%%%%%%%%%%%%%%%%%%%%%%%%%%%%%%%%%%%%%%%%%%%%%%%%%%%%%%%%%%%%%%%%%%%%%%%%%%%%%%%
%% Keywords should appear after the \end{abstract} command. 
%% See the online documentation for the full list of available subject
%% keywords and the rules for their use.
\keywords{protoplanetary disks --- polarization 
--- ISM: jets and outflows ---} 
%%%%%%%%%%%%%%%%%%%%%%%%%%%%%%%%%%%%%%%%%%%%%%%%%%%%%%%%%%%%%%%%%%%%%%%%%%%%%%%%%

%% From the front matter, we move on to the body of the paper.
%% Sections are demarcated by \section and \subsection, respectively.
%% Observe the use of the LaTeX \label
%% command after the \subsection to give a symbolic KEY to the
%% subsection for cross-referencing in a \ref command.
%% You can use LaTeX's \ref and \label commands to keep track of
%% cross-references to sections, equations, tables, and figures.
%% That way, if you change the order of any elements, LaTeX will
%% automatically renumber them.

%% We recommend that authors also use the natbib \citep
%% and \citet commands to identify citations.  The citations are
%% tied to the reference list via symbolic KEYs. The KEY corresponds
%% to the KEY in the \bibitem in the reference list below. 

%%%%%%%%%%%%%%%%%%%%%%%%%%%%%%%%%%%%%%%%%%%%%%%%%%%%%%%%%%%%%%%%%%%%%%%%%%
\section{Introduction} \label{sec:intro}
%%%%%%%%%%%%%%%%%%%%%%%%%%%%%%%%%%%%%%%%%%%%%%%%%%%%%%%%%%%%%%%%%%%%%%%%%

The study of protoplanetary disks has recently seen 
significant advances, driven by the desire to identify the initial conditions for planet formation. The advent of the Atacama Large Millimeter/submillimeter Array (ALMA) has brought increased sensitivity in mm-wave polarimetry, opening a new area of investigation into disk properties. 
Polarimetry is long believed to allow access to information on the orientation of magnetic field lines, since linear polarization can arise from 'grain alignment', i.e. the tendency of non-spherical dust grains to align their short axis along the magnetic field lines (e.g. \citealt{Andersson2015}). 
This is a crucial test for disk models, in particular regarding the magneto-centrifugal acceleration of outflows \citep{BP82, Frank2014} and the magneto-rotational instability \citep{Balbus1991}.
However, linear polarization in the mm continuum emitted from disks can also be produced by processes unrelated to the magnetic field. In particular, self-scattering and radiative grain alignment can result in a high percentage of linear polarization, with very specific polarization patterns \citep{Kataoka2017, Yang2017, Tazaki2017}.
%However, linear polarization can also arise through self-scattering of thermal emission of dust grains, with size comparable to $\lambda/2\pi$, where $\lambda$ is the radiation wavelength \citep{Yang2017, Kataoka2015, Kataoka2017}. Models indicate that the polarization induced by this mechanism in inclined disks would bepredominantly oriented parallel to the minor axis.  
%In addition, polarization can also arise from alignment of non-spherical dust grains with their short axis parallel to the direction of an anisotropic radiation field. 
%For a centrally illuminated disk, the linear polarization would present for this mechanism  a circular pattern centred on the source (e.g. \citet{Tazaki2017}). 
%\citep{Lazarian-Hoang2007, Zeng2013, Tazaki2017}. 

The first polarimetry studies of protostellar envelopes, at 1$^{\prime\prime}$-3$''$ resolution, allowed the identification of hourglass-shaped, twisted polarization patterns consistent with the pinching of magnetic field lines due to the contraction of the natal cloud \citep{Girart2006, Rao2009, Hull2014}. Subsequent studies at 0$\farcs$4 - 1$\farcs$2 reported the first detections of polarized emission in protostellar disks (e.g., \citealt{Rao2014, Cox2015, Kataoka2016a}). 
%In none of these cases, however, the polarization structure was easily interpretable in terms of the expected magnetic field configuration. 
%who observed a spiral polarization pattern 
%consistent with a toroidal field  in the face-on disk of IRAS 16293B and in .... 
%The angular resolution available for these studies, however, 
%5was not high enough to reveal polarization in the disk itself. 
%A first result in this drection was obtained by  Rao et al. 2014, who observed a spiral polarization pattern consistent with a toroidal 
%field  in the face-on disk of IRAS 16293B. 
Recently, the scales of the inner protoplanetary disk have been accessed with ALMA, reaching 0$\farcs$1 resolution. 
%and  sensitivity to detect polarization fraction down to  0.5 - 1\%. 
Observations show that all the mechanisms mentioned above can produce 
polarization, but  dust self-scattering appears to be dominant here \citep{Stephens2014, Kataoka2016b, Stephens2017, Kataoka2017, Lee2018, Girart2018, Hull2018, Alves2018}.

Our study maps the polarization properties of the more evolved 
(i.e. Class II) systems which are associated with jets.  
These systems offer the advantage of being  
%, to try and relate the properties of disks and jets. 
less embedded than younger sources. In addition, the kinematics of the bipolar jet allows identification of the disk near-side, which is an important information in polarization studies \citep{Yang2017}.  
%and the assumed 
%magneto-centrifugal origin for the jets offers 
%constraints on the expected magnetic configuration.
Here, we present the results for two sources, CW Tau and DG Tau, selected because they are nearby (d$\sim$140 pc, \citealt{Rebull2004}), tilted with respect to the plane of the sky (i.e. favorable for polarization measurements), and have bright dust continuum emission (allowing sensitivity in detection of polarization). 
Previous studies of these sources include disk investigations 
%at mm wavelengths 
by \citet{Testi2002, Isella2010, Pietu2014} and jet investigations by \cite{Eisloeffel1998, Bacciotti2002, Coffey2007,  Dougados2000, Hartigan2004}. 

 %%%%%%%%%%%%%%%%%%%%%%%%%%%%%%%%%%%%%%%%%%%%%%%%%%%%%%%%%%%%%%%%%%
\begin{figure*}
%\begin{figure}
 \gridline{\fig{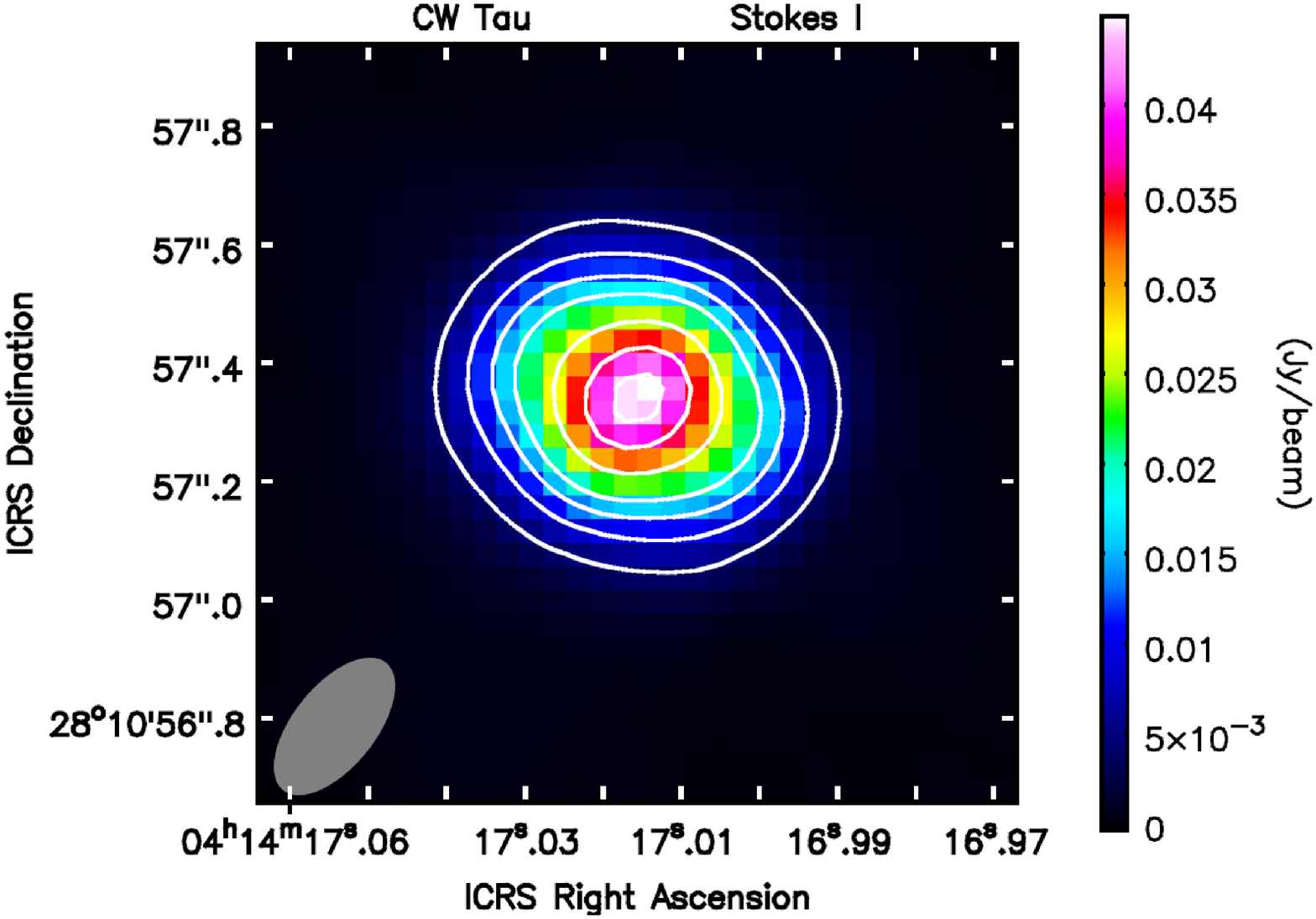}{0.5\textwidth}{(a)}
          \fig{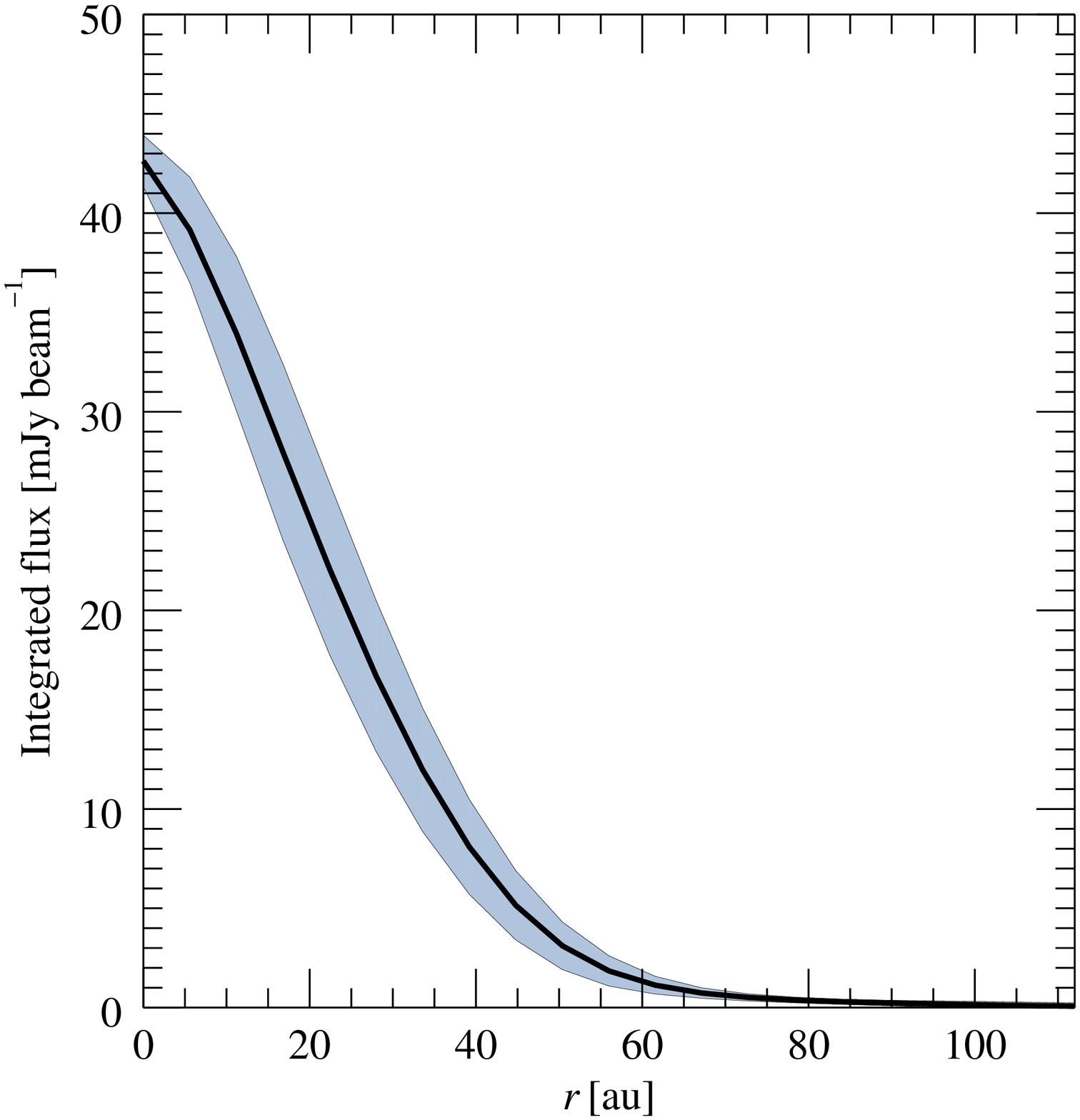}{0.35\textwidth}{(b)}
          }
\gridline{\fig{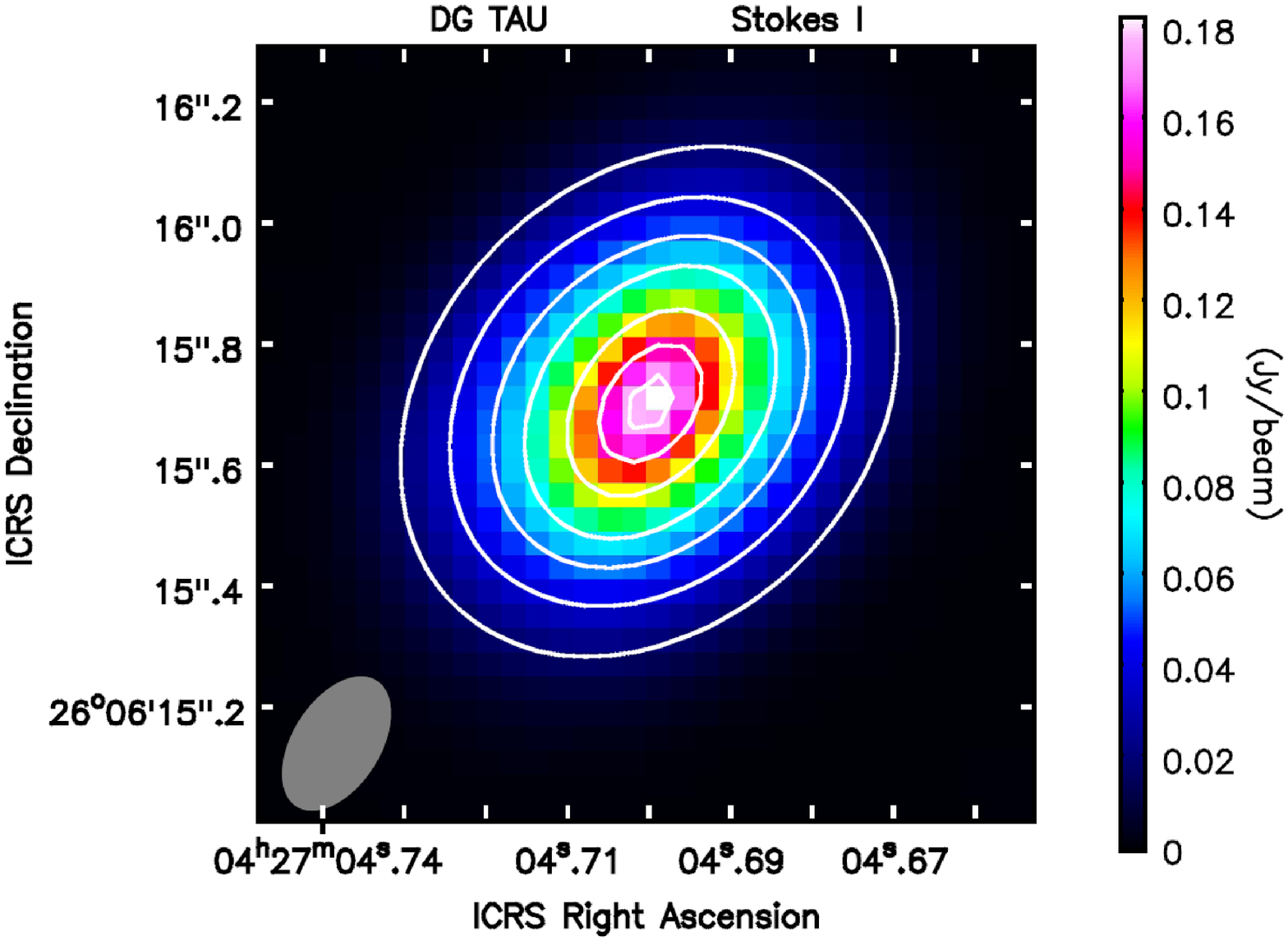}{0.5\textwidth}{(c)}
          \fig{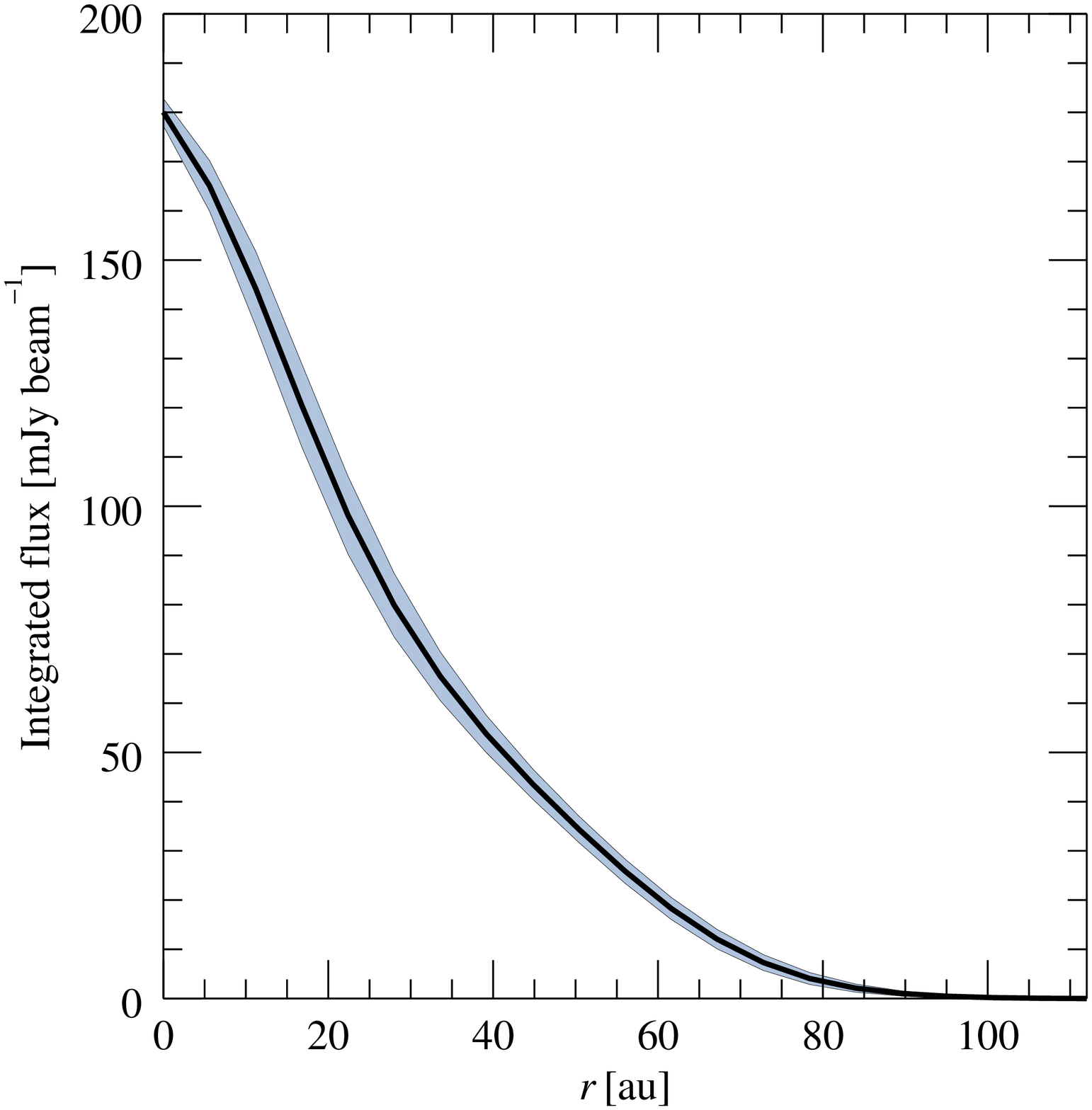}{0.35\textwidth}{(d)}
          }                  
%----------------------------------------------------------------
%\includegraphics[width=\columnwidth]{CWTAU_StokesI_v3.eps}
%\includegraphics[width=\columnwidth]{DGTAU_StokesI_v3.eps}
%----------------------------------------------------------------
%\plottwo{DGTAU_POLRATIO_JET.eps}{DGTAU_POLRATIO_JET.eps}
%\includegraphics[height=6.7cm]{DGTAU_STokesI_v3.eps}
%\includegraphics[height=6.7cm]{CWTAU_StokesI_v3.eps}
%\plotone{DGTAU_POLRATIO_JET.eps}
\caption{Total emission map at 870 $\mu$m and average radial intensity profiles 
in the disks around CW Tau  (panels (a),(b)) and DG Tau (panels (c),(d)).
Contour levels are [0.1,~0.2,~0.3,~0.4,~0.6,~0.8,~0.95] $\times$ peak value, 
which is 44.9 and 182.4 mJy beam$^{-1}$ for CW Tau and DG Tau respectively.
The grey areas in the radial profile give the 
1$\sigma$ uncertainties from the standard deviation. 
%(after deprojection for the disk inclination and ) 
%The ellipse in the lower left corner is the synthesized beam 
%(0$\farcs$27$\times$0$\farcs$14, PA=$-39^{\circ}$, for CW Tau 
%and 0$\farcs$24$\times$0$\farcs$14, PA=$-32^{\circ}$ for DG Tau).
\label{fig:mapsI}}
%\end{figure}
\end{figure*}
%%%%%%%%%%%%%%%%%%%%%%%%%%%%%%%%%%%%%%%%%%%%%%%%%%%%%%%%%%%%%%%%%%%  

 %%%%%%%%%%%%%%%%%%%%%%%%%%%%%%%%%%%%%%%%%%%%%%%%%%%%%%%%%%%%%%%%%%
\begin{figure*} 
%\begin{figure}
\gridline{\fig{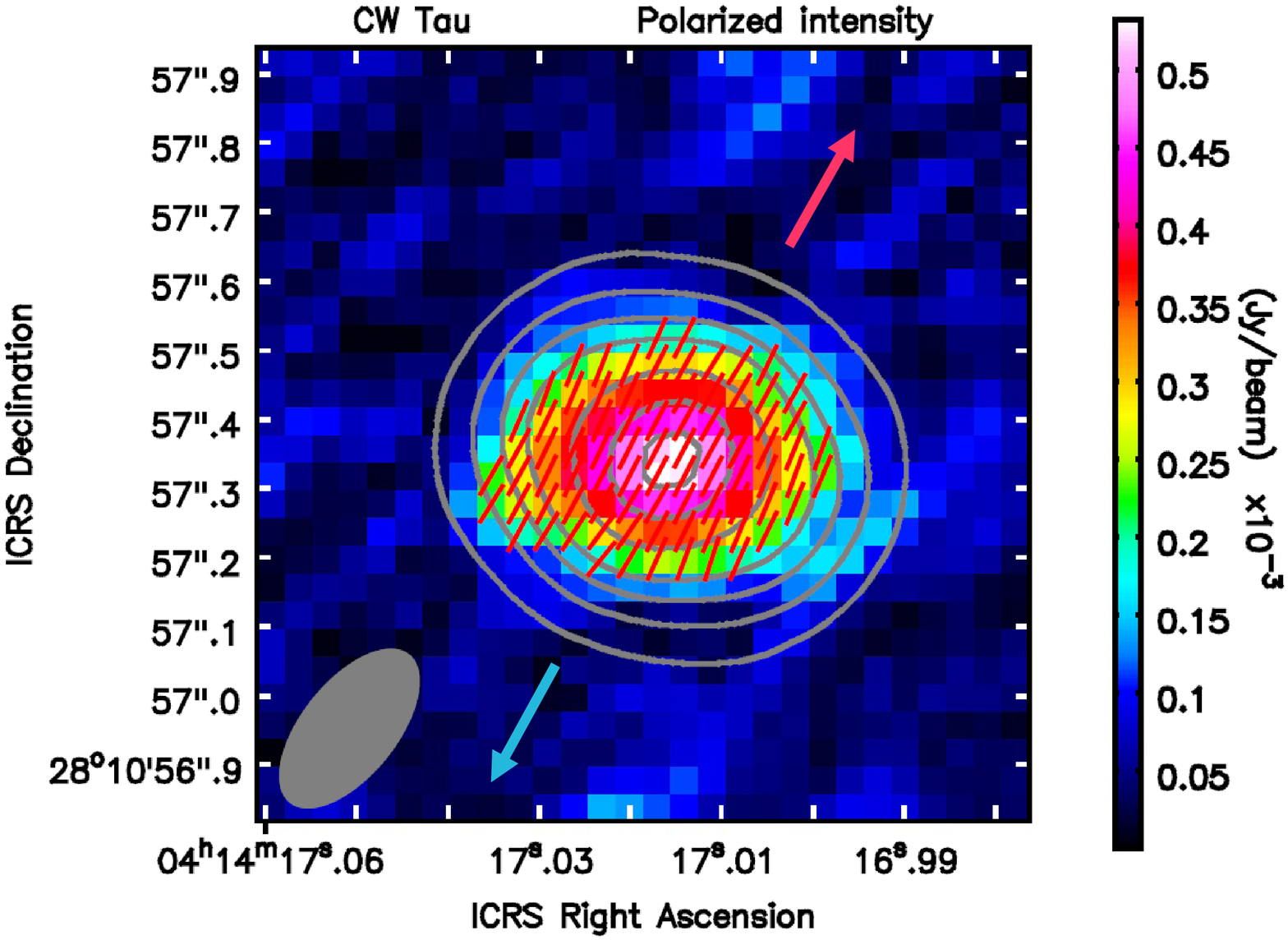}{0.5\textwidth}{(a)}
          \fig{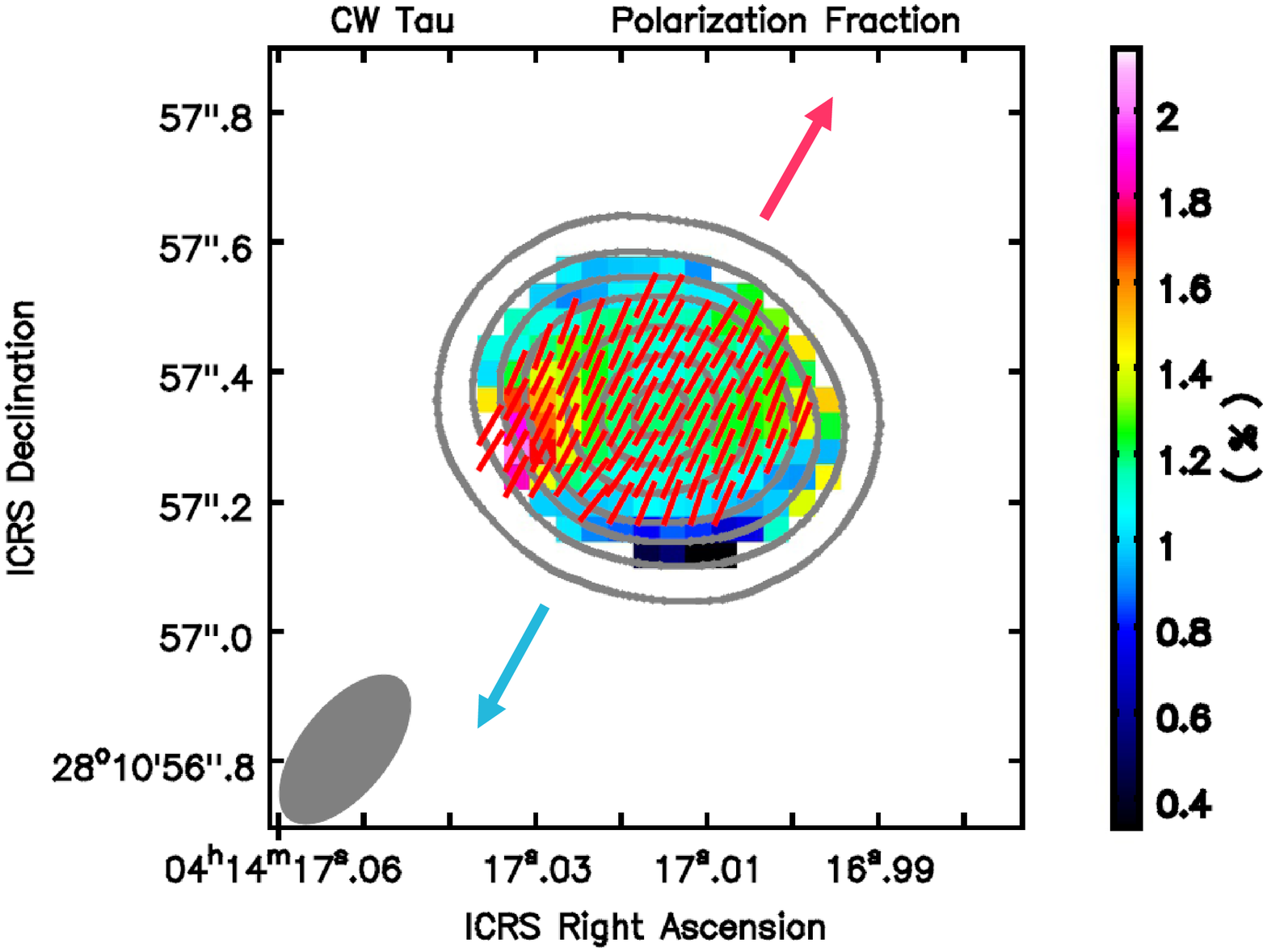}{0.5\textwidth}{(b)}
          }
\gridline{\fig{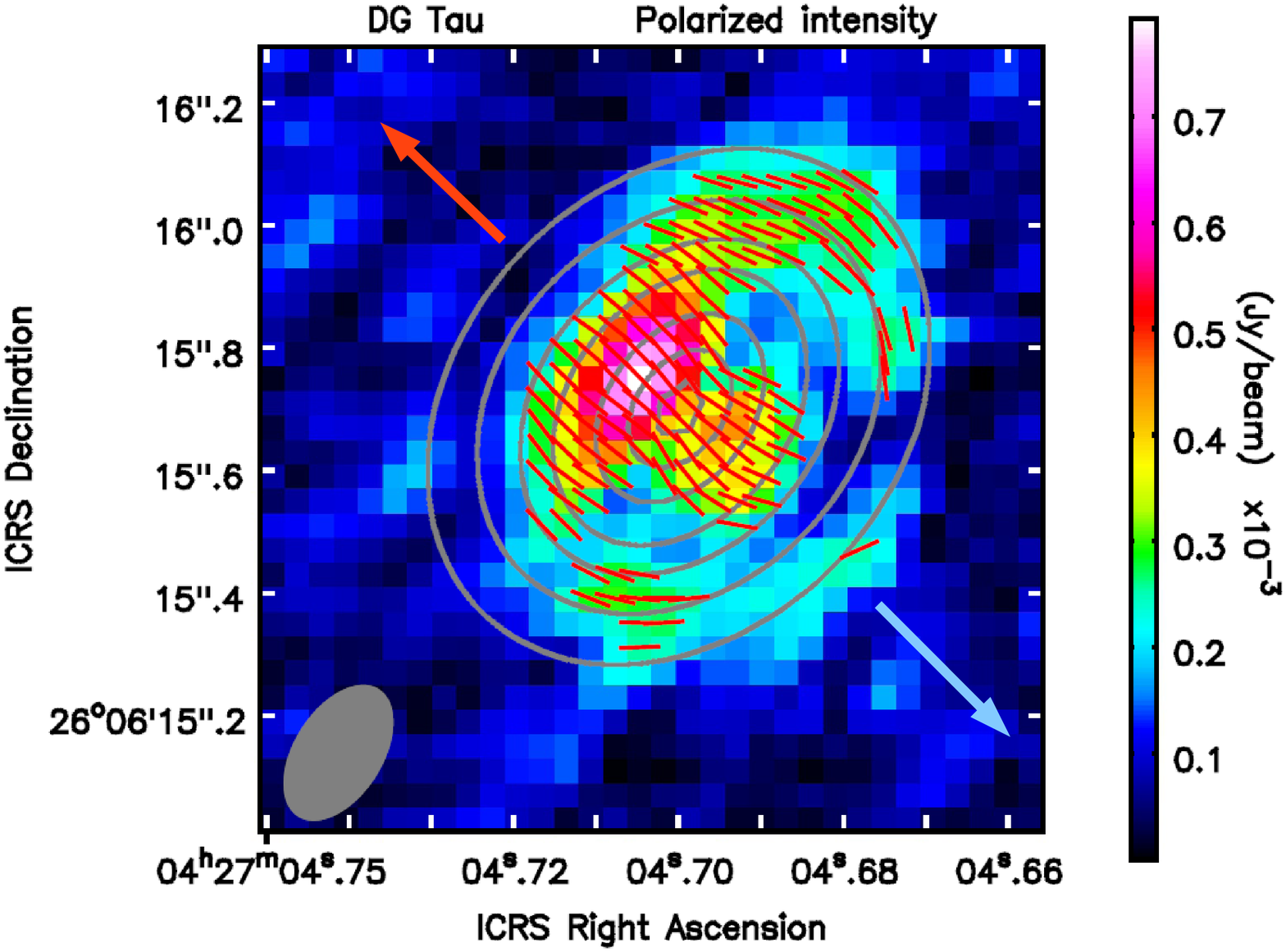}{0.5\textwidth}{(c)}
          \fig{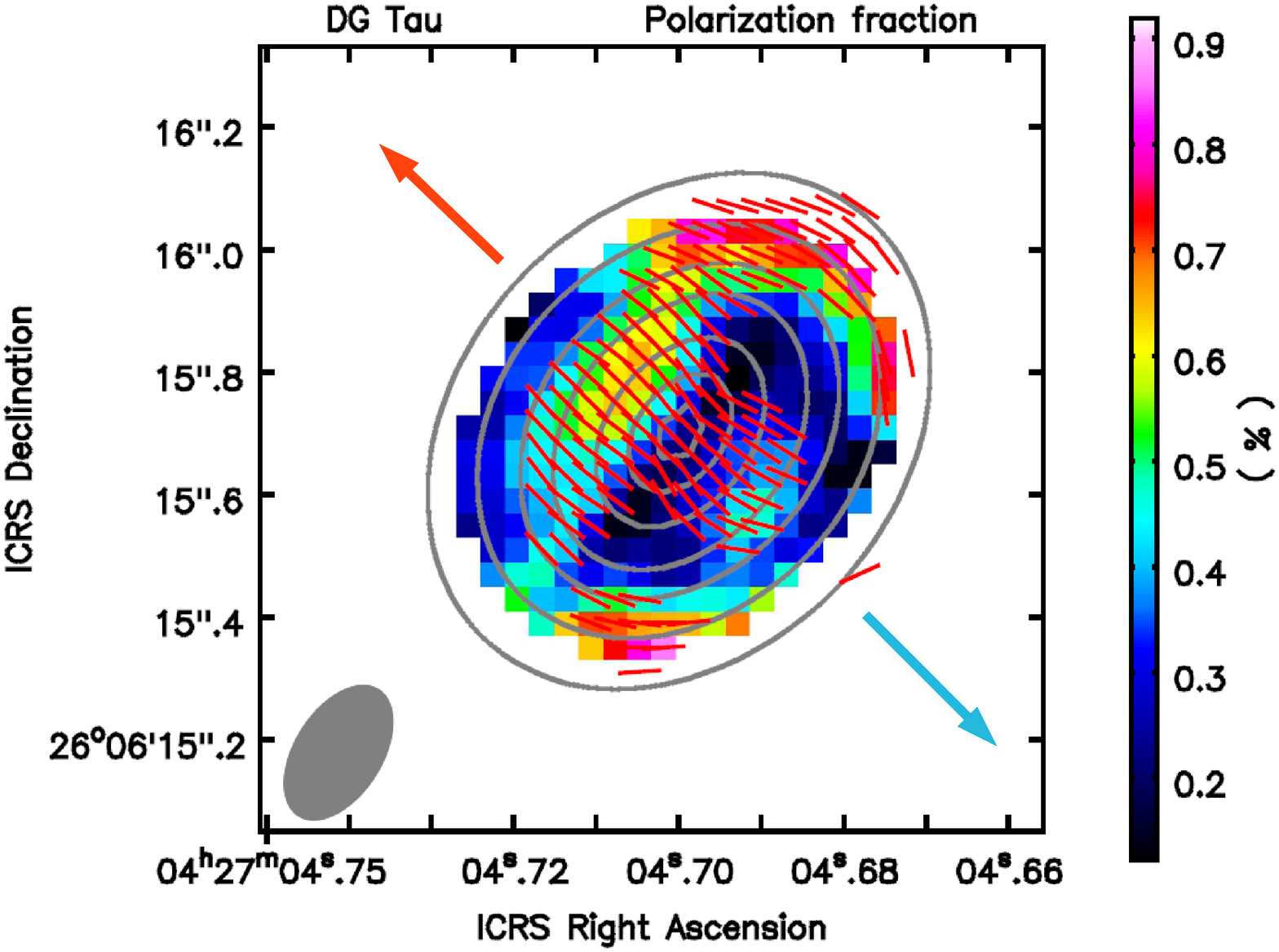}{0.5\textwidth}{(d)}
          }
%----------------------------------------------------------------
%\includegraphics[width=\columnwidth]{CWTAU_StokesI_v3.eps}
%\includegraphics[width=\columnwidth]{DGTAU_StokesI_v3.eps}
%----------------------------------------------------------------
%\plottwo{DGTAU_POLRATIO_JET.eps}{DGTAU_POLRATIO_JET.eps}
%\includegraphics[height=6.7cm]{DGTAU_STokesI_v3.eps}
%\includegraphics[height=6.7cm]{CWTAU_StokesI_v3.eps}
%\plotone{DGTAU_POLRATIO_JET.eps}
\caption{Linearly polarized intensity $P$  and  polarization fraction $p$ 
at 870 $\mu$m in the disks around CW Tau (panels (a),(b)), and DG Tau (panels (c),(d)).  
Contours as in Fig. \ref{fig:mapsI}.
Polarization angle, $\chi$, is indicated with fixed-length vector bars.  
Arrows indicate the jet orientation.  
Disk near-side lies towards the receding jet lobe (red arrow). Polarization fraction
$p$ is shown where total intensity is  $I>$ 10 mJy beam$^{-1}$  for CW Tau and $I>$ 30 mJy beam$^{-1}$ 
for DG Tau.  
\label{fig:mapsP}}
%\end{figure}
\end{figure*}
%%%%%%%%%%%%%%%%%%%%%%%%%%%%%%%%%%%%%%%%%%%%%%%%%%%%%%%%%%%%%%%%%%%  

%%%%%%%%%%%%%%%%%%%%%%%%%%%%%%%%%%%%%%%%%%%%%%%%%%%%%%%%%%%%%%%%%%%%%%%%%%
\section{Observations and data reduction} \label{sec:obs}
%%%%%%%%%%%%%%%%%%%%%%%%%%%%%%%%%%%%%%%%%%%%%%%%%%%%%%%%%%%%%%%%%%%%%%%%%%

We observed polarized emission towards the young Taurus systems 
CW Tau 
%(simbad RA 4h 27m 4.7s,  $\delta$ 28$^{\circ}$ 6' 16.0")
and DG Tau, 
%(simbad RA 4h 14m 17.0s, $\delta$ 28$^{\circ}$ 10' 57.4")
within the ALMA Cycle 3 program 2015.1.00840.S (PI: F. Bacciotti). 
Observations were carried out in Band 7 (870 $\mu$m) in full polarization mode. 
The spectral setup included four spectral windows, 1.875 GHz wide, centred at 
the standard ALMA band 7 polarization frequencies (336, 338, 348 and 350 GHz).
The spectral resolution was 31.250 MHz (55 km s$^{-1}$).
Two successful executions 
%as part of the session scheme 
were made on 2017 July 11, 
with 40 antennas in the configuration C36-6, giving 
nominal angular resolution of 0$\farcs$15.
Total exposure time was 34.38 min for CW Tau and 30.24 min for DG Tau.
%During the session the q
Quasars J0429+2724 and J0403+2600 were observed to
calibrate the bandpass, and the gains in amplitude and phase,
respectively. 
To determine the instrumental contribution to the cross-polarized
interferometer response, a bright strongly polarized (2\%) quasar, J0522-3627, was observed. 

The data were processed using the Common Astronomical Software
Application (CASA) version 4.7.2. 
The two datasets were calibrated separately to obtain total intensity, and 
then concatenated to perform the polarization calibration. Calibration followed the 
standard procedure described in \citet{Nagai2016} and in the ALMA
polarization casaguide\footnote{https://casaguides.nrao.edu/index.php/3C286\_Band6Pol\\
\_Calibration\_for\_CASA\_4.3\#Polarization\_Calibration}. 

The data were reduced using the task {\tt\string CLEAN} in CASA. To improve the  dynamic range of the images, phase-only self-calibration runs were executed. For CW Tau, 
two runs resulted in a final dynamic range
of 625, a factor of 3 better than in the initial image. Meanwhile, for DG Tau, three runs resulted in a final dynamic range of 1025, giving an improvement of a factor 7
with respect to the initial image. 
The final images were obtained using Briggs weighting, and the
synthesized beam was 
0$\farcs$27$\times$0$\farcs$14 (PA=$-39^{\circ}$) for CW Tau and
0$\farcs$24$\times$0$\farcs$14 (PA=$-32^{\circ}$) for DG Tau,
(corresponding to an average of 29 au and 27 au, respectively, at 140 pc). 
%facendo la media aritmetica come in stephen 2017
The r.m.s. achieved for 
CW Tau was 72 $\mu$Jy beam$^{-1}$
in Stokes I, and $\sim$ 40 $\mu$Jy beam$^{-1}$ in Stokes Q and U, while
for DG Tau it was 180 $\mu$Jy beam$^{-1}$ in Stokes I, and $\sim$
58 $\mu$Jy beam$^{-1}$ in Stokes Q and U.
From the Stokes I, U, Q maps we obtain the linear polarization intensity, 
$P=\sqrt{Q^2 + U^2 }$, the linear polarization fraction, $p=P/I$, 
and the polarization angle, $\chi=0.5 \arctan (U/Q)$, 
i.e. the direction of polarization of the electric field.  
The ALMA instrumental error is reported to 
be  0.1\% on $p$, and at least $2^{\circ}$ on $\chi$ 
(ALMA polarization casaguide). 
\vspace{0.5cm}

%%%%%%%%%%%%%%%%%%%%%%%%%%%%%%%%%%%%%%%%%%%%%%%%%%%%%%%%%%%%%%%%%%%%%%%%%%
\section{Results} \label{sec:res}
%%%%%%%%%%%%%%%%%%%%%%%%%%%%%%%%%%%%%%%%%%%%%%%%%%%%%%%%%%%%%%%%%%%%%%%%%%

\subsection{Total Emission}
Figure~\ref{fig:mapsI} illustrates the total intensity (Stokes I) of the 870 $\mu$m continuum emission and the azimuthal average of the intensity radial profile for each target, with both showing a smooth distribution with no features apparent at our resolution. 

For CW Tau, the integrated flux is 145.1 $\pm$ 1.4 mJy,                          % 0.14508 con IMSTAT 139.4 with IMFIT - IMSTAT RMS 0.61
with peak intensity of 44.9 $\pm$ 0.3 mJy beam$^{-1}$                         %44.97 con IMSTAT 
located at RA 4h 14m 17.0s, $\delta$ 28$^{\circ}$ 10$^{\prime}$ 57$\farcs$35.           %28$^{\circ}$ 10' 57$\farcs$34 imfit.
%ra: 04:14:17.015514 +/- 0.000100 s (0.001321 arcsec along great circle)
%dec: +028.10.57.343067 +/- 0.001071 arcsec
The measured flux is the same, within the errors, as the one
% 35  * (1.3/0.87)^(2.3) * (0.27*0.14)/(0.6*0.4) (linear scaling: peaked emission) = 13 but if 0.15/0.3 = 44 
% 58.7 \pm 0.4  * (1.3/0.87)^(2.3)  = 147 pm 2
extrapolated from integrated 1.3 mm flux of \citet{Pietu2014}, adopting their spectral index of 2.3. 
The FWHM along the major and minor axis, determined with a 2D Gaussian fit 
deconvolved from the beam, is  0$\farcs$35 and 0$\farcs$18 respectively. 
From these values, we estimate a disk inclination $i_{\rm disk}$ with respect to the line of sight of $\sim 59^{\circ}$, 
%exactly
between the inclination angle for the disk, $i_{\rm disk}= 65 \pm 2^{\circ}$ indicated by  \cite{Pietu2014} 
and the one of jet, $i_{\rm jet} = 49^{\circ}$ reported in \cite{Hartigan2004}. 
The same fit gives a disk position angle, PA$_{\rm disk}$ = 60.7$^{\circ} \pm {1.9^\circ}$, 
in good agreement with PA$_{\rm disk}=62^{\circ} \pm 3^{\circ}$ \citep{Pietu2014}, and almost 
perpendicular to the jet PA$_{\rm jet}=$-$29^{\circ}$ \citep{Hartigan2004}.

For DG Tau, the integrated flux is 880.2 $\pm$ 9.4 mJy,                      
% 0.880 con IMSTAT 864.2 com IMFIT 
with peak intensity of 182.4 $\pm$ 1.4 mJy beam$^{-1}$,                    
%182.37 with IMSTAT 152.6 con IMFIT -  IMSTAT RMS 15.12
 located at %(simbad RA 4h 27m 4.7s,  $\delta$ 28$^{\circ}$ 6' 16.0")
RA 4h 27m 4.7s, $\delta$ 26$^{\circ}$ 6$^{\prime}$ 15$\farcs$71.          % 26$^{\circ}$ 6' 15$\farcs$71 imstat
%ra: 04:27:04.69979 +/- 0.00013 s (0.00178 arcsec along great circle)
%dec: +026.06.15.70328 +/- 0.00188 arcsec
The integrated flux is slightly lower (by 12\%) than that extrapolated from 1.3 mm flux of \citet{Isella2010}, adopting their spectral index of 2.5. 
% 49 pm 4.5 (5sigma)  * (1.3/0.87)^(2.5) * (0.24*0.14)/(0.15*0.17) (linear scaling: peaked emission) = 143 pm 20  
% 367 pm 14  * (1.3/0.87)^(2.5)  = 1001 pm 50 
A 2D Gaussian fit deconvolved from the beam provided 
a FWHM along the major and minor axis of 0$\farcs$45 and  0$\farcs$36, respectively. 
The combination of these values implies $i_{\rm disk} \sim 37^{\circ}$, 
%arccos(36/45)=37 
almost identical to $i_{\rm jet} = 38^{\circ}$ \citep{Eisloeffel1998}, 
and only slightly higher than disk models of $i_{\rm disk}= 24^{\circ} $--$ 32^{\circ}$ \citep{Isella2010}. 
The 2D Gaussian fit gives PA$_{\rm disk}$ = 135.4$^{\circ} \pm {2.5^\circ}$, almost perpendicular to PA$_{\rm jet}$ = 46$^{\circ}$ \citep{Eisloeffel1998}. 

\subsection{Polarized Emission}
Figure~\ref{fig:mapsP} illustrates the polarization properties 
in our targets.
%of CW Tau (panels (a), (b)) and of DG Tau (panels (c), (d). 
%Here we refer to the direction of polarization of the electric field.
%The color map in the top panel is the intensity of polarized emission $P=\sqrt{Q^2+U^2}$ in mJy beam$^{-1}$, 
%while the  map in the bottom panel is the polarization fraction $p$ in percent. 
%(The high values measured at the borders of the $p$ maps are due to $I$ dropping to zero, and so should be taken with caution.) 
%The polarization angle $\chi$ is indicated by the direction of vector bars (of fixed length). 
%The length of the vectors is constant and independent on the polarization intensity. 
%In all the panels, the contours refer to the total intensity as in Fig.\ref{fig:f1}. The blue and red arrows indicate the direction of the approaching and receding lobes of the jet. 

\begin{figure}
%----------------------------------------------------------------
\includegraphics[width=\columnwidth]{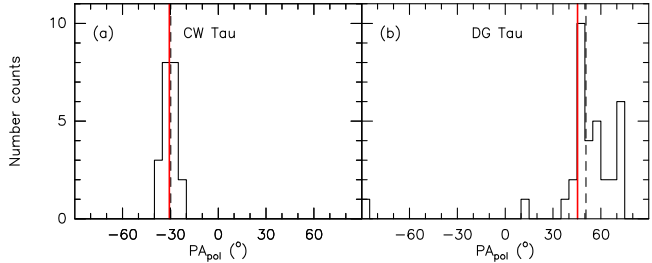}
%----------------------------------------------------------------
\caption{Distribution of the 
polarization angles toward CW Tau (a) and DG Tau (b). Dashed and red lines indicate 
the average of the distributions and the PA of the disk minor axis, respectively. 
\label{fig:PAhist}}
%\end{center}
\end{figure}

For CW Tau, the map of linearly polarized intensity $P$ is centrally peaked and does not show any significant asymmetry. The maximum value is 0.53 $\pm$ 0.14  mJy beam$^{-1}$. 
The superimposed polarization vectors 
%(parallel to the radiation electric field) 
are aligned along the minor axis of the disk. 
Figure \ref{fig:PAhist}, panel (a), 
shows that the distribution  of the polarization angles
is very narrow, averaging $-29.8^{\circ} \pm 4.0^{\circ}$,  
and nearly coincident with the PA of the disk minor axis and the PA of the jet. The polarization fraction, $p$, is  almost constant in the central region of the disk, 
%0$\farcs$3$\times$0$\farcs$3 region. 
averaging 1.15 $\pm$ 0.26 \% over the whole disk. 

%%%%%%%%%%%%%%%%%%%%%%%%%%%%%%%%%%%%%%%%%%%%%%%%%%%%%%%%%%%%%%%%%%
%%%%%%%%%%%%%%%%%%%%%%%%%%%%%%%%%%%%%%%%%%%%%%%%%%%%%%%%%%%%%%%%%%%

For DG Tau, the intensity map is distributed asymmetrically with respect to the disk's major axis. The near-side of the disk (identified by its proximity to the receding jet lobe) is brighter, with the signal distributed parallel to the major axis.  The peak emission of 0.79 $\pm$ 0.13 mJy beam$^{-1}$ is found along the minor axis, displaced by $\sim 0\farcs$07 from the Stokes I photocenter. 
A secondary peak of 0.45 mJy beam$^{-1}$ is seen  
0$\farcs$07 south-west of the disk centre.
In the outer disk region, between 0$\farcs$3
and 0$\farcs$5 from the source, the polarized emission is distributed
in a belt-like structure of lower intensity (0.2-0.3 mJy beam$^{-1}$).  
The polarization vectors
%, visible in Fig.\ref{fig:mapsP} as superimposed red segments, 
follow two distinct patterns. In the central inner region, they appear to be nearly aligned with the minor axis, while in the outer belt they have an azimuthal orientation. 
The distribution of polarization angles (Fig.~\ref{fig:PAhist}, panel (b)) gives an average of 50.7$^{\circ}$ $\pm 27.4^{\circ}$, which is similar to the PA of the minor axis (45$^{\circ}$). 
The wide distribution is due to the azimuthal orientation of the vectors in the outer disk. 
%The transition between the two regions is discussed in Section~\ref{transition}. 
The linear polarization fraction $p$ reflects the distribution of the polarized intensity, with a region of higher values in the disk's near-side, along a direction parallel to the major axis. 
The peak of 0.65 $\pm$ 0.10\% is located  at 0$\farcs$07 from the total intensity peak. The outer belt at 0$\farcs$5 
shows a higher polarization fraction toward the NW and the SE,
where  the polarized emission falls off more slowly than the total intensity. 
The average value of the polarization fraction  
over the whole disk area is 0.41 $\pm$ 0.17 \%, with a median of 0.38\%.

%%%%%%%%%%%%%%%%%%%%%%%%%%%%%%%%%%%%%%%%%%%%%%%%%%%%%%%%%%%%%%%%%%%%%%%%%%
\section{Discussion} \label{sec:disc}
%%%%%%%%%%%%%%%%%%%%%%%%%%%%%%%%%%%%%%%%%%%%%%%%%%%%%%%%%%%%%%%%%%%%%%%%%%

\subsection{Polarization from Self-Scattered Dust Emission} 

The polarization properties in our sources match the model expectations of self-scattering of thermal dust emission 
from large ($>$30 $\mu$m) dust grains, as modelled by \cite{Kataoka2015, Kataoka2017},
and \cite{Yang2016, Yang2017}, and in agreement with the findings in other protoplanetary disks at similar wavelengths (e.g.
\citet{Kataoka2017} for HL Tau,  \citet{Girart2018} for the  GGD27 MM1 disk, \citet{Hull2018} for the disk around IM Lup). 
In all these cases, the strongest indication for polarization from self-scattering 
is the alignment of the polarization vectors along the minor axis in the central region of the disk. 
This feature originates from the geometry of the scattering 
when the disk is inclined with respect to the line of sight.
%Furthermore, the observed values of the polarization fraction are 
%in the range  predicted by the models.
%as it is the increase at the outer borders of the disk. 

\subsection{Grain Size Estimates} 
\citet{Kataoka2015} show that the maximum  grain size contributing to polarization from self-scattering 
at a given wavelength, $\lambda$, is comparable to $\lambda/2\pi$. At 870 $\mu$m, 
we expect a maximum size in the range 35 -- 350 $\mu$m, peaked around 140 $\mu$m (see their Fig. 3). 
Further constraints can come from the correlation between  grain size, wavelength and polarization fraction, 
as investigated for the disk around HL Tau in \citet{Kataoka2016b} and \citet{Kataoka2017}.  
We attempt a rough estimate of the grain size in our targets using the same diagnostic diagrams, 
because the polarization fraction has a weak dependence on the particular disk model  
while it is strongly dependent on the grain size
\citep{Kataoka2016b}. 
Using our values of the average polarization fraction and the diagrams in  
\citet{Kataoka2017}  we estimate that 
the maximum grain size giving rise to the observed polarization is  
in the range 100 - 150 $\mu$m for CW Tau, and in the range 50 - 70 $\mu$m for DG Tau.

\subsection{Constraints on Dust Settling}
In CW Tau, the distribution of the polarized intensity is symmetric, and polarization vectors are nearly parallel to the 
minor axis, with no curvature toward the outer  disk.
These features are consistent with polarization produced by self-scattering in either an optically 
thin disk \citep{Yang2016}, or in an optically thick but geometrically thin disk 
(\cite{Yang2017}, see their Fig.\,10). 
\citet{Pietu2014} 
%$\beta \sim 1.3$, 
%$\alpha = 2.25$. Given $\alpha < 2 + \beta$,  
find that this dusty disk is optically thick, thus our observations indicate that the grains in question 
are located in a geometrically thin layer near the disk midplane. 

Meanwhile, in DG Tau, we see a similar polarization angle alignment in the inner disk but in this case it is accompanied 
by an asymmetry in polarization intensity. This combination is consistent with the expectations of models of self-scattering in disks 
of intermediate or high optical depth, and with a finite angular thickness \citep{Yang2017}.  
%%%%%%%%%%%%%%%%%%%%%%%%%
We estimate the optical depth of the DG Tau disk at 870 $\mu m$ 
using the models for dust opacity and surface density 
in \citet{Isella2010}. 
The optical depth varies smoothly from about 1.0 to  about 0.4 
between 10 and 30 au from the star, where we observe the asymmetry, 
and decreases to about 0.1 at 80 au
(an uncertainty of 20\% accounts for the different models adopted).
The disk is, therefore, moderately optically thick, which corresponds to the assumptions of Model B 
of \cite{Yang2017}. Indeed, the observed polarization map is well reproduced by the output  
of this model (see Figures 4d, 4e, 4f in \cite{Yang2017}). 
The observed asymmetry appears thus to  
indicate that the scattering 
grains have not yet settled to the midplane.

%%%%%%%%%%%%%%%%%%%%%%%%%%%%%%%%%%%%%%%%%%%%%%%%%%%%%%%%%%%%%%%%%%
\begin{figure}
%\begin{center}
%----------------------------------------------------------------
\includegraphics[width=\columnwidth]{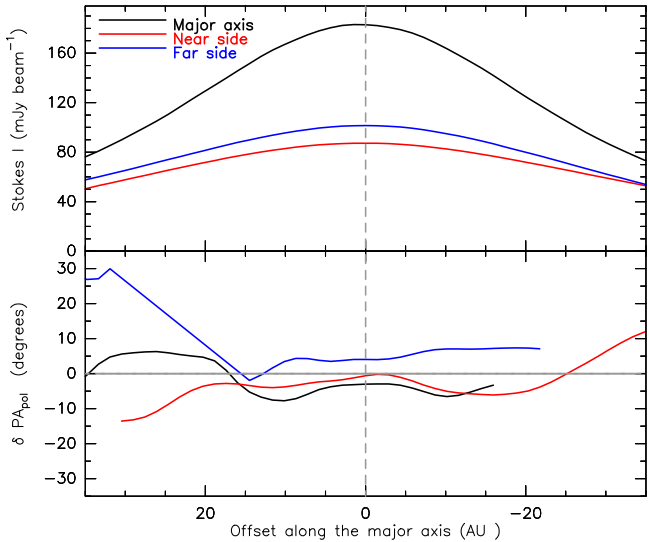}
%----------------------------------------------------------------
\caption{ 
{\em Upper panel}: total intensity profiles across the DG Tau disk along cuts parallel to the major axis 
- black line represents the major axis itself (PA$_{\rm disk}$ = 135$^{\circ}$) - 
red line cuts the disk at 0$\farcs$14 
%($\sim 20$ au on the plane of the sky) 
%north of the major axis 
to its near side - blue line cuts at 0$\farcs$14 %south of the major axis (
to its far side. 
Positive offsets correspond to east of the intensity peak.
{\em Lower Panel}: difference between the position angles of the polarization vectors and 
of the disk minor axis, along the same profile cuts as above.   
\label{fig:3cuts}}
%\end{center}
\end{figure}
%%%%%%%%%%%%%%%%%%%%%%%%%%%%%%%%%%%%%%%%%%%%%%%%%%%%%%%%%%%%%%%%%%%

%%%%%%%%%%%%%%%%%%%%%%%%%%%%%%%%%%%%%%%%%%%%%%%%%%%%%%%%%%%%%%%%%%
\begin{figure*}
%\begin{center}
\gridline{\fig{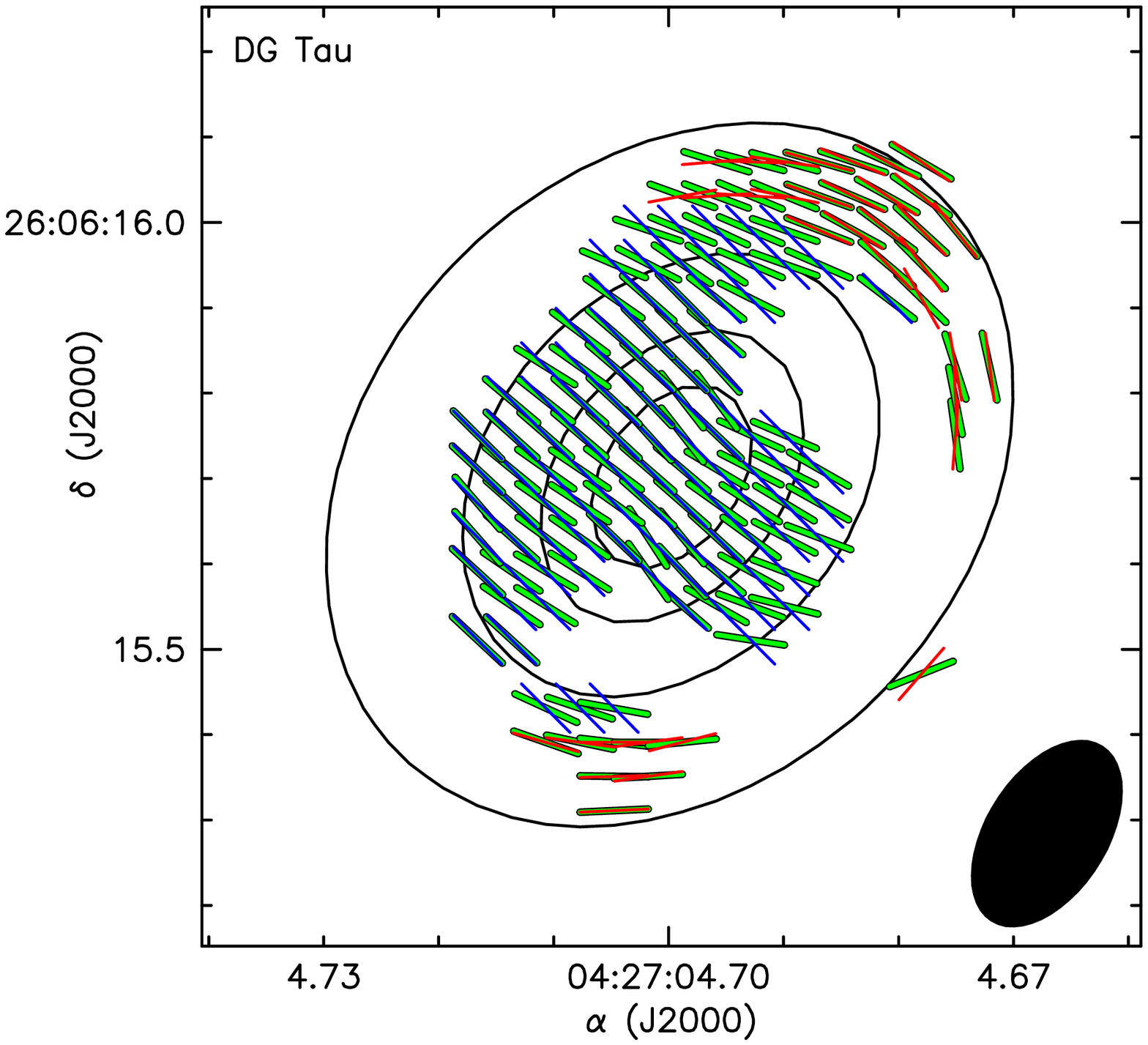}{0.5\textwidth}{(a)}
          \fig{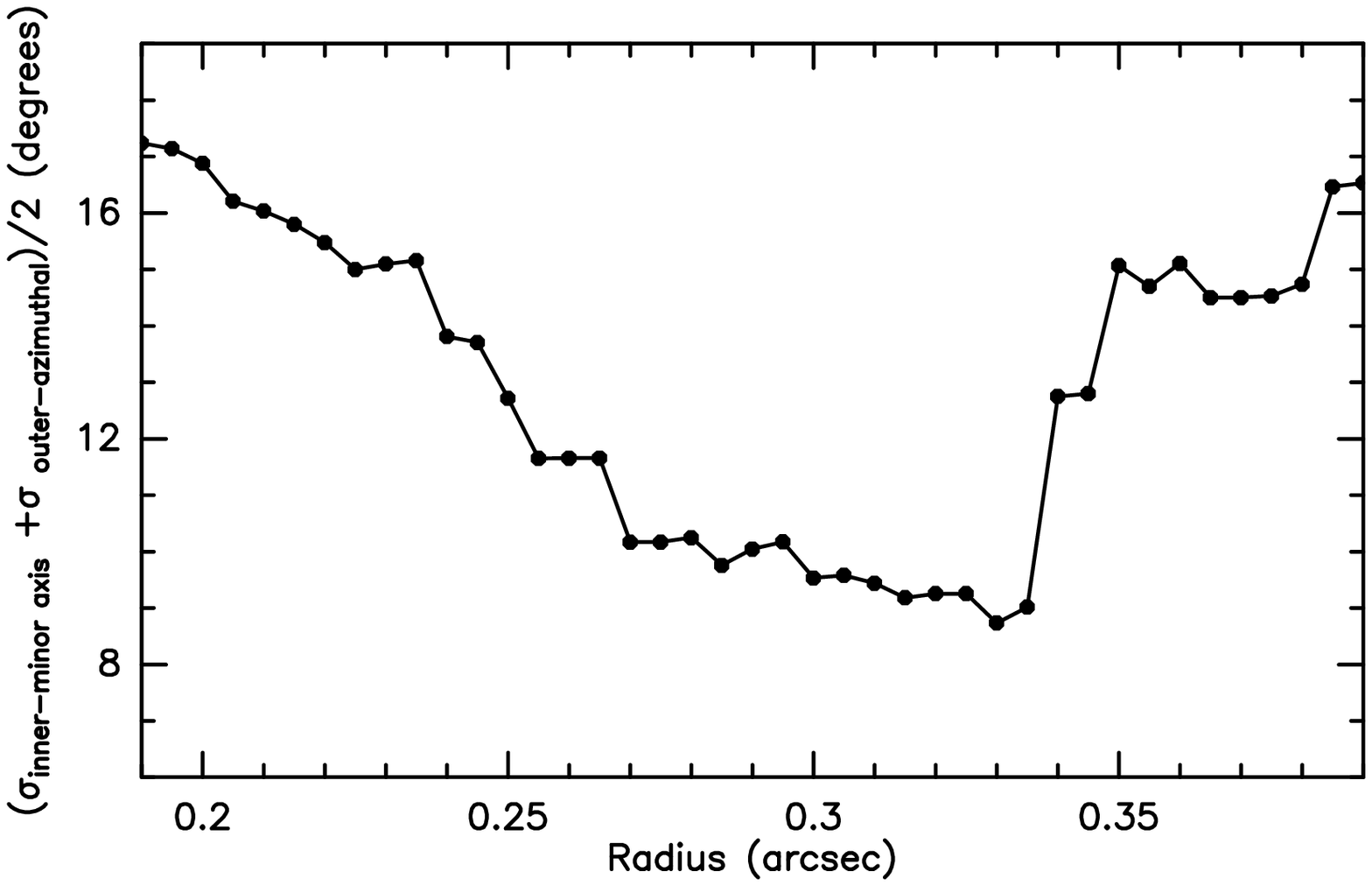}{0.4\textwidth}{(b)}
          }
%\gridline{\fig{dgprofile.eps}{0.5\textwidth}{(c)}}
%----------------------------------------------------------------
\caption{ 
Panel (a): observed polarization angles (green), with superposed a model polarization pattern 
in which the vectors lie parallel to the minor axis (blue) for radii smaller than  a given $r=R$,  
while at larger radii the vectors lie along the azimuthal angle (red). 
Here  $R=0\farcs33$ has been adopted, as the value that minimizes the combined standard 
deviation (panel (b))  of the differences between the observed polarization angles and those in the  model.  
Contours in panel (a) represent total intensity, from 20 to 180 mJy beam$^{-1}$ in steps of 40 mJy beam$^{-1}$.
\label{fig:PAmodel}}
%\end{center}
\end{figure*}
%%%%%%%%%%%%%%%%%%%%%%%%%%%%%%%%%%%%%%%%%%%%%%%%%%%%%%%%%%%%%%%%%%%

We further investigate DG Tau by comparing the PA of the polarization vectors within the inner 35 au ($\sim 0$\farcs$25$) to that of the disk minor axis. 
Figure \ref{fig:3cuts} (lower panel) shows PA differences along three cuts parallel to the major axis:
the major axis itself (black); 0$\farcs$14 towards the near-side (red); and 0$\farcs$14 towards the far-side (blue).
The PA differences are less than 10 degress in the inner 20 au, 
and do not change sign  at the minor axis. 
This is in contrast with the expectations from 
the models at moderate/high optical thickness in \cite{Yang2017}, in which the PA difference is expected 
to change sign crossing the minor axis, and to reach an absolute value between 15 and 20 degrees.   
%In this case, the asymmetry in polarized intensity is less pronounced but still present, and 
%there is a residual low-level polarized emission  in the disk far-side,
%along the minor axis, as we observe in DG Tau. The bifurcation in the polarization position angles 
%in the inner disk is  reduced, and the  polarization vectors in the outer disk 
%tend to align along the azimuthal direction (see Figure 4f in  \cite{Yang2017}).
This suggests that in the DG Tau disk the grains are distributed 
with a finite angular thickness (because we observe an asymmetry), but the optical depth of this material is not high enough to induce the position angle bifurcation. 
It is  possible that our observations probe  an intermediate scale height between 
the disk surface and the disk midplane. Alternatively, we may be witnessing moderate dust settling occurring in  
an optically thick portion of the disk.

\subsection{Variation of Polarization Orientation in DG Tau}
\label{transition} 
We now consider the outer region of the DG Tau disk, i.e. beyond 0$\farcs$3  from the source. Figure \ref{fig:mapsP}, panel (c), shows structures in the polarized emission which do not correspond to any feature in the total intensity at the same resolution. In addition, a change in the orientation of the polarization pattern is observed. 

This PA transition was investigated using a geometrical model in which the polarization vectors are parallel to the minor axis inside a given radius $R$, and have an azimuthal orientation outside $R$, as in Figure~\ref{fig:PAmodel}, panel (a). 
The Figure illustrates the best fit with the observations, obtained setting $R=0\farcs33$.
This value of $R$ is the one for which the 
the standard deviation in PA difference (combining inner and outer disk) as a function of (the free parameter) $R$ reaches its minimum (see
Figure~\ref{fig:PAmodel}, panel (b)). The good agreement with the observed polarization pattern suggests that there is a real change in polarization properties at about 45 au from the star. 
The physical meaning of this radius, however, which has been derived under purely geometrical assumptions, is not obvious. 
No substructure is apparent in the total intensity map, nor in the averaged radial profile of the total intensity, at our resolution (Fig. \ref{fig:mapsI}, panels (c,d)). 
A possible explanation may come from a drop in the optical depth at $R$. This would imply that the outer disk is optically thin and that the radiation is seen to come primarily from the inner disk. Under such conditions, the self-scattered radiation  would have an azimuthal polarization orientation in the outer disk, as observed.
A similar pattern is found by \cite{Girart2018} in the  GGD27 MM1 disk, 
and a change in optical depth is invoked for the transition.
The physical reason for such a change, however, remains to be investigated.  
Alternatively, alignment of non-spherical grains with the anisotropic radiation field \citep{Tazaki2017} 
may contribute to the azimuthal polarization pattern in the outer disk belt. 
Multi-wavelength observations at higher angular resolution are needed to clarify this point.
In any case, our analysis indicates that the polarization maps are providing us with complementary information on the disk structure which are not accessible via the total intensity maps.

\section{Conclusions}
Our ALMA observations of the disks around CW Tau and DG Tau support dust self-scattering as the origin of the polarization seen in the 870 $\mu$m continuum emission.
By observing systems with jets, we can successfully identify the near-side of the disk and thus correctly interpret asymmetry properties of the linearly polarized light intensity. 
This in turn provides us with constraints on the scale height 
of the dust grains responsible for this polarization. 
Furthermore, our determination of the polarization fraction allows us to
estimate a maximum size for the scattering dust grains. 
We find that such grains tend to be larger 
and more settled to the disk midplane in the case of CW Tau than in DG Tau. 
The maps of the polarization intensity and polarization vector orientation  
reveal details not  shown in the total intensity 
map at the same resolution. 
%%, which provide independent information on the disk properties.
%We are consequently motivated to further investigate these systems with multi-wavelength observations at similar and higher angular resolution, as well as with more detailed modeling. 
%% of the polarized emission. 
%% In summary, 
Overall, our analysis indicates that polarimetry is set to be a powerful tool in accessing information on the dust structure, 
which cannot be obtained from the unpolarized emission alone. 
It is anticipated that this will be of great interest in studies of structure and evolution of protoplanetary disks. 

%%%%%%%%%%%%%%%%%%%%%%%%%%%%%%%%%%%%%%%%%%%%%%%%%%%%%%%%%%%%%%%%%%%%%%%%%% 
\acknowledgements
%%%%%%%%%%%%%%%%%%%%%%%%%%%%%%%%%%%%%%%%%%%%%%%%%%%%%%%%%%%%%%%%%%%%%%%%%% 
This paper uses ALMA data from project ADS / JAO.ALMA 2015.1.00840.S. 
ALMA is a partnership of ESO (representing its member states), NSF (USA) and NINS (Japan), together with NRC (Canada), MOST and ASIAA (Taiwan), and KASI (Republic of Korea), in cooperation with the Republic of Chile. 
The Joint ALMA Observatory is operated by ESO, AUI/NRAO and NAOJ.  
We thank the anonymous referee for very valuable comments. 
JMG is supported by the MINECO (Spain) AYA2014-57369-C3 
and AYA2017-84390-C2 grants. MP acknowledges funding from the EU-Horizon-2020/MSC grant agreement No-664931.
LP and CC acknowledge  the project PRIN-INAF/2016 GENESIS-SKA. 
This work was partly supported by the 
Italian Ministero dell\'\,Istruzione, Universit\`a e Ricerca, through the grants
Progetti Premiali  2012/iALMA (CUP-C52I13000140001), 2017/FRONTIERA (CUP-C61I15000000001), SIR-(RBSI14ZRHR) (acknowledged by DF and CF), and by
the Deutsche Forschungs-gemeinschaft (DFG, German Research Foundation)-Ref no. FOR 2634/1 TE 1024/1-1, and by the DFG cluster of excellence Origin and Structure of the Universe (\href{http://www.universe-cluster.de}{www.universe-cluster.de}).

%%%%%%%%%%%%%%%%%%%%%%%%%%%%%%%%%%%%%%%%%%%%%%%%%%%%%%%%%%%%%%%%%%%%%%%%%%

% \citep  (Io et al. 2000)
% \citep{io, tu }  (io, tu) 
%(e.g. % \citealt )  (e.g. Io et al. 2000)  (e.g. \citealt{io, tu })  ( e.g. io, tu)
% \cite, \citet    Io et al. (2000)

%%%%%%%%%%%%%%%%%%%%%%%%%%%%%%%%%%%%%%%%%%%%%%%%%%%%%%%%%%%%%%%%%%%%%%%%%%

\end{document}